\documentclass[final]{svjour2}
\usepackage{graphicx}
\usepackage{rotating}
\usepackage{amssymb}
\usepackage{mathptmx}
\usepackage[numbers]{natbib}
\makeatletter
\journalname{Journal of Low Temperature Physics}

\bibpunct{}{}{,}{s}{}{,}

\begin{document}

\newcommand{\hdblarrow}{H\makebox[0.9ex][l]{$\downdownarrows$}-}

\title{Study of the dependency on magnetic field and bias voltage of
  an AC-biased TES microcalorimeter.}

\author{L. Gottardi$^1$ \and J. Adams$^2$ \and C. Bailey$^2$ \and
  S. Bandler$^2$ \and M. Bruijn$^1$ \and J. Chervenak$^2$ \and M. Eckart$^2$ \and F. Finkbeiner$^2$ \and R. den Hartog$^1$ \and
  H. Hoevers$^1$ \and R. Kelley$^2$ \and C. Kilbourne$^2$ \and P. de Korte$^1$ \and J. van der Kuur$^1$ \and M. Lindeman$^1$ \and F. Porter$^2$ \and J. Sadlier  $^2$ \and S. Smith$^2$}

\institute{1:SRON National Institute for Space Research,\\ 
Sorbonnelaan 2, 3584 CA Utrecht, The Netherlands\\
\\2: NASA GSFC,\\
Greenbelt Road, Greenbelt, MD 20771, USA}

\date{XX.XX.2007}

\maketitle


\begin{abstract}
At SRON we are studying  the performance of a Goddard Space Flight Center
single  pixel   TES  microcalorimeter  operated  in   an  AC  bias
configuration. For  x-ray photons  at 6 keV  the pixel shows  an x-ray
energy resolution $\Delta E_{FWHM}=3.7$eV, which is about a factor 2 worse than
the  energy resolution observed  in an  identical DC-biased  pixel. In
order to
better understand the reasons for this discrepancy we characterised the
detector as a function of  temperature, bias working point and applied
perpendicular magnetic field.  A strong periodic dependency of the
detector noise on the TES AC  bias voltage is measured. We discuss the
results in  the framework of the recently  observed weak-link behaviour
of a TES microcalorimeter.

PACS numbers: 
\end{abstract}

\section{Introduction}

In previous works we compared the performances of an x-ray TES
microcalorimeter under AC and DC bias by measuring the IV
characteristic, the noise, the impedance and the x-ray response at
perpendicular magnetic field $B_\perp=0$. The tests were carried out
both with an SRON pixel and a GSFC pixel \cite{GottardiAC}. With
respect to the DC  bias case, under AC bias we observe  a smaller
(about $15 \%$) current output at small voltage bias low in the
transition, a low TES current and temperature sensitivity , a
slightly worse integrated NEP resolution and about a factor two x-ray
resolution degradation.       
In order to better understand the reasons for the suboptimal
performance of the TESs under AC bias  we
thoroughly studied the effect of the perpendicular magnetic field and
the voltage bias on
the detector response. Here we present the results obtained with a GSFC
pixel read out both in the  AC and DC configuration.

\section{Experimental details}

\begin{figure}[htbp] 
\centering                                                             
\includegraphics[width=0.5\textwidth,keepaspectratio]{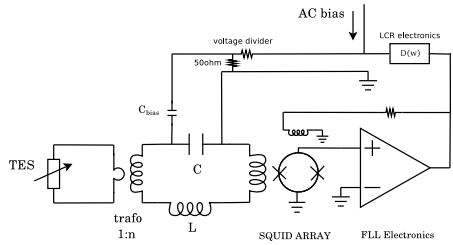}
\caption{Schematic drawing of the AC bias and read-out circuit used for the TES microcalorimeter. A superconducting flux transformer is  used to improve the impedance matching between the
SQUID amplifier and the TES microcalorimeter. \label{ACreadout2}}
\end{figure}

A  schematic   drawings  of  the   read-out  circuit used for the
AC measurements described here is   shown  in
Fig.~\ref{ACreadout2}. A superconducting flux transformer is used between the SQUID
amplifier and the TES microcalorimeter to  improve the  impedance
matching between the TES and SQUID amplifier.  

The  TES microcalorimeter   is tested using a  superconducting  transformer with
inductances   $L_p=100\, \mathrm{nH}$   and   $L_s=6.4 \, \mu\mathrm{H}$,   and
estimated    mutual    inductance   $M=k^2L_sN_p/N_s=760 \, \mathrm{nH}$,
$N_{r}=N_s/N_p=8$ coil  turns ratio and  with $L_p$ connected  to the
TES.   The  impedance of  the  LC  circuit seen  by  the  TES is  then
$Z_{LC,tes}=Z_{LC}/N_r^2$.   The LC  resonator consists  of  an hybrid
filter with  a lithographic Nb-film  coil with $L<10\, \mathrm{nH}$  and commercial
high-Q NP0  SMD capacitors with  $C=100\pm10 \,\mathrm{nF}$.  The circuit
has    an     additive    total    stray     inductance    of    about
$L_{stray}\sim 200\,\mathrm{nH}$. The intrinsic resonator
factor of the LC resonator  is $350\pm 20$, limited by losses in the lumped elements circuit.

As amplifiers  we  used a NIST  SQUID arrays consisting of a  series of 100  dc-SQUID with
 input-feedback  coil  turns  ratio  of  3:1,  and  input  inductance
 $L_{in}=70 \, \mathrm{nH}$.   The   input   current   noise   is   $\sim
 4 \, \mathrm{pA}/\sqrt{Hz}$ at  $T<1\, \mathrm{K}$.  The SQUID  amplifier is
 operated  in a  standard analogue  flux-locked-loop (FLL)  mode using
 commercial Magnicon electronics, which linearises the
 SQUID response.  Only at  frequencies below $f<700 \, \mathrm{kHz}$ the FLL
 has   enough  loop   gain  to   sufficiently  linearise   the  signal
 response.  For  this reason  our  experiments  are  carried out at  a
 resonant frequency of $470 \, \mathrm{kHz}$.

The circuit   resonant  frequency is defined by the  capacitor C  and  the  total  inductance
 $L_{tot}=L_{in}+L+L_{stray}+L_{fb}+L_{s}-M_{ps}^2/(L_p)$, where the last
term accounts for the screening effect of the superconducting
transformer and  $L_{fb}$ is an inductance generated by the SQUID 
feedback loop. From the measured resonant frequency and
the filter capacitance reported above we get $L_{tot}=1.1\,\mu
\mathrm{H}$.  The effective inductance seen by the TES is $L_{eff,TES}=1.1 \, \mu \mathrm{H}/{N_r}^2= 17 \mathrm{nH}$.

In the experiment described here we used an xray TES microcalorimeter from
the GSFC.  It is  a $150\times 150 \mathrm{\mu m^2}$  pixel  from  a  uniform  $8\times8$  array
\cite{GSFCarray2008},  where  a  TES  is  coupled  to  a  micron-thick
overhanging   Au/Bi   X-ray   absorber.   The  sensor   is   a   Mo/Au
proximity-effect bilayer  with a transition temperature of  $T_C=95\,
\mathrm{mK}$, and  a  normal state  resistance  of  $R_N=7\,\mathrm{m}\Omega$.   

\section{Experimental results}

\begin{figure}[h]                                                              
 \center
 \includegraphics[width=1.\textwidth,angle=0]{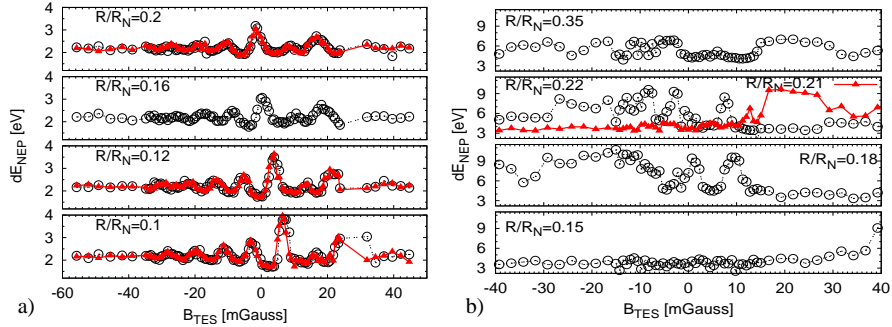}\caption{
   Integrated NEP as a function of magnetic field and for several bias
   point with TES DC biased ({\bf a.}) and AC biased ({\bf b.}). The
   cancelling magnetic field for the DC and AC bias pixels are
   respectively $-228$ and $68$ mGauss.  The
   points in red up triangles  are the results of measurements taken
   under identical experimental condition, but on a different day.\label{NEP_ACDCbias}}
\end{figure}

When the GSFC pixel is DC biased good baseline and  x-ray energy
resolution is observed. The x-ray resolution is comparable to the
baseline resolution and is generally of the order of
2.3-2.5eV. 
The pixel responsivity and noise strongly depends on  the
perpendicular magnetic field applied to the TES. In
Fig.\ref{NEP_ACDCbias}{\bf a}.  we show the
integrate NEP ($dE_{NEP}$) as a function of the applied magnetic field $B$
and for different bias current. At this pixel
we observed a remanent perpendicular field of $B=228 \mathrm{mGauss}$.
The pattern observed  is due to the
dependency of the detector critical current on the  magnetic
field as a result of the TES behaving as a weak-link \cite{Sadlier2010}.
The shift of the Josephson patterns along the applied magnetic field
is due to the self magnetic field generated
by the DC current flowing through the leads connecting the TES
\cite{SmithASC2010}.

Under AC bias we measured  the $dE_{NEP}$ as a function
of the applied perpendicular magnetic field $B$ as well. 
At this pixel
we observed a remanent perpendicular field of $B=-68\mathrm{mGauss}$.
The results are shown in Fig.~\ref{NEP_ACDCbias}{\bf b}. 

From the plots we observe that, under AC bias: the  Fraunhofer-like
oscillations are visible, but the  pattern is more noisy and less
reproducible than under DC bias; the baseline resolution is slightly
worse and strongly depends on the bias voltage.
While the former effect is likely due to the self magnetic field, the latter has 
a less trivial explanation. 
In an attempt to clarify these results, we performed a fine scan
of the detector I-V characteristic by measuring for every bias point
the detector x-ray response and the noise. We repeated the scan for
different magnetic fields. 
We discovered that the baseline energy resolution strongly depends on the detector bias point. A very small variation
($\sim 2\%$) of the bias current  could result in an energy resolution
degradation of more than a factor three. 

\begin{figure}[h]
 \center
 \includegraphics[width=1.\textwidth,angle=0]{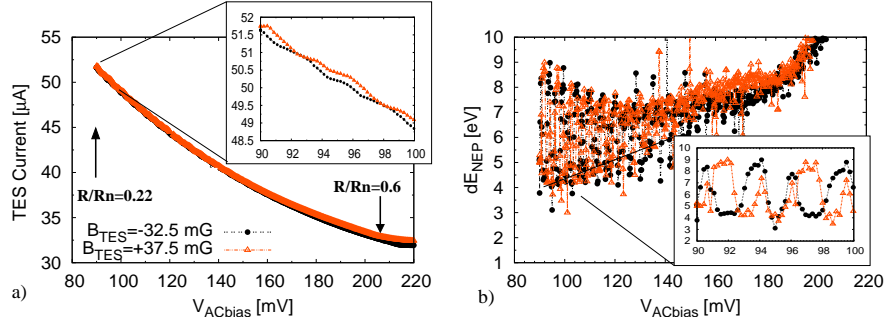}\caption{ \label{dESIvsIbias}
   Signal amplitude (i.e. $I_{rms,TES}$ TES current) and integrated
   energy resolution as a function of the bias voltage of the AC
   biased pixel. The
 energy resolution shows an oscillating pattern  as a function of  the
 bias point. The pattern is partially modulated by the magnetic
 field. $T_{bath}=65$mK.}
\end{figure}

The results of this measurement are shown
in Fig.\ref{dESIvsIbias}, where the baseline resolution
is plotted as a function of the bias voltage and for several value of the perpendicular magnetic field.  

The energy resolution oscillates between values from 3.5 eV to 9 eV as a function of  the bias point.  
The oscillating pattern is partially modulated by the magnetic
 field. 

As visible in the insert of Fig.\ref{dESIvsIbias}{\bf a.}, the
sensor IV curve presents a staircase structure, modulated by the
perpendicular magnetic field. 

\begin{figure}
 \center
 \includegraphics[width=0.45\textwidth,angle=0]{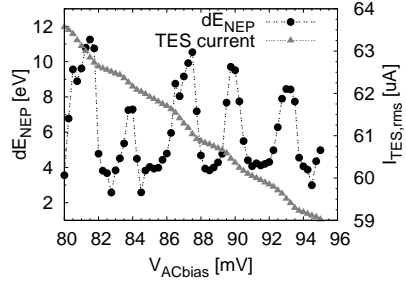}\caption{ \label{dEvsIbiaszoom}
   Integrated energy resolution as a function of the bias voltage. The
 energy resolution deteriorates at the bias point corresponding
 to the higher slope in the IV curve steps. $T_{bath}=18$ mK.}
\end{figure}

The worst resolution is measured at the transition between two
 steps where the slope of the I-V curve is the highest. This
is clearly seen in Fig.\ref{dEvsIbiaszoom}. 

\begin{figure}
 \center
 \includegraphics[width=1.\textwidth,angle=0]{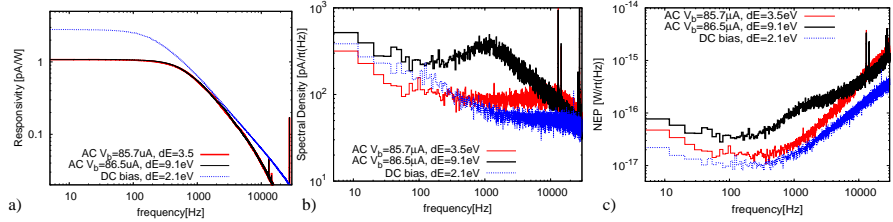}\caption{
   Responsivity ({\bf a.}), noise ({\bf b.}) and NEP ({\bf c.}) measured under DC bias and AC bias. For the AC bias case the
   responsivity for two  bias points is shown, taken respectively at the flat (red curve) and
   at the rising (black curve) part of the step.\label{SiNoiseACDC}}
\end{figure}

In Fig.~\ref{SiNoiseACDC} we plot the NEP,
the responsivity and the noise spectra taken under AC and DC bias. 
For the AC bias case the spectra are taken respectively at the flat and
   at the rising part of the observed steps in Fig.~\ref{dEvsIbiaszoom}.
The detector response bandwidth is not identical under AC and DC bias due
to the different load impedance of the two circuits
\cite{GottardiAC}. One should refer to the NEP plot when comparing the
AC and DC cases.
The NEP is the lowest under DC bias. At low
frequency ($f<100 \mathrm{Hz}$) the degradation in the NEP
  observed in the AC bias case is probably due to a
  reduced responsivity.  At high frequency ($f>1\mathrm{kHz}$) excess
  noise is observed in the AC bias case. The excess noise is worse for
  the measurement taken at the rising part of the steps and has the
  signature of excess Johnson noise.
  Furthermore, the noise level at frequency $f>1\mathrm{kHz}$  in the spectra
  corresponding to the 3.5 eV  integrated NEP taken under AC bias,
  cannot be explained by simply including  the SQUID and the LC
  resonator thermal noise. An excess white noise of
  about $50\mathrm{pA}/\sqrt{Hz}$ is estimated from the model. 
Under AC bias the responsivity is independent on the two bias points.  

\section{Discussion}

In analogy with the analysis done for an rf-SQUID
\cite{Jackel1975,Kurkij,SilverZimmer67,ClarkeBook} we calculate the
characteristic parameters of a TES in a superconducting loop  weakly
coupled via a superconducting transformer to an LC resonant
circuit. The TES is treated  as a weak-link in accordance with  the RSJ
model where $R_{shunt}$ is assumed to be TES normal resistance.
For the detector described above we find the cut-off and the characteristic
frequency to be respectively $\omega_{cut}=R/L\sim
400\mathrm{kHz},\;     \;    \omega_{JJ}=2\pi RI_c/\Phi_0\sim 100 \mathrm{MHz}.$
\noindent where a critical current of about $I_c\sim 5\mu\mathrm{A}$ is
  assumed. Note that the TES critical current at a $T_{bath}\ll T_c$
  is generally larger than $5\mu\mathrm{A}$ ($700\mu\mathrm{A}$ at
  55mK \cite{SmithASC2010}). 
However in bias conditions the TES operating temperature is  $T\sim T_c$ and the critical  current drops.  
The screening parameter is $\beta_{rf}=2\pi LI_0/\Phi_0\sim 273$. 
The rfSQUID-like AC biased TES operates then in an adiabatic and
hysteretic regime, since  $\omega_{LC}< min(\omega_{JJ},\omega_{cut})$ and
$\beta_{rf}\gg 1$. rf-SQUIDs have shown the
highest noise in this working regime \cite{ClarkeBook}.

The steps in the TES IV curves have a non zero slope (Fig.\ref{dESIvsIbias}{\bf a.}).  The tilting of the steps
in the rf-SQUID I-V characteristic and the rounding of the step edges
reflect directly the width of the quantum transitions distribution due to
thermal fluctuations. Kurkij\"arvi \cite{Kurkij}
have shown that the ratio $\alpha$ of the voltage rise along a step $\Delta V_s$ to
the voltage difference $\Delta V_0$ between steps is directly
proportional to the SQUID's instrinsic flux noise \cite{Jackel1975}.
Their  empirical formula gives
$\alpha=\frac{1}{0.7\Phi_0}\big{(}\frac{\omega_{rf}}{2\pi}\big{)}^{1/2}\sqrt{S_{\Phi}}.$
In the same way we can estimate the
flux noise in the TES superconducting ring. From the step observed in
the IV curves we get an $\alpha=\Delta
 V_s/\Delta V_0=0.36$, which corresponds to a flux noise of
 $\sqrt{S_{\Phi}}\sim 4\cdot 10^{-4}\mathrm{ \frac{\Phi_0}{\sqrt{Hz}}}$.

\noindent For a 17 nH inductance ring this is equivalent to  a current
noise of $\sqrt{S_{I}}\sim 5\cdot 10^{-11}\mathrm {\frac{A}{\sqrt{Hz}}}$.
This noise level is comparable with the excess noise observed in the
AC bias detector, which limits the best measured  integrated NEP to
$3.5 \mathrm{eV}$.

\section{Conclusions}

We observed excess noise and low reproducibility in the AC biased
x-ray pixel. 
A strong dependency of the baseline noise on the bias voltage is at the origin of the non reproducible
results obtained in the past.  
In the IV curve under AC bias a staircase structure is observed. 
A possible interpretation of this effect is given by comparing the
detector and the AC read-out to an rf-SQUID. In this analogy the TES
is seen as a weak-link in a superconducting ring weakly coupled to an
LC resonator. Should this interpretation be valid,  the excess noise
observed in the AC bias experiment could be due to flux noise
generated by the uncertainties in the quantum transition. 

To validate this hypothesis the following tests should be performed in
the future: measurements with large L and high bias frequency ($f>1 \mathrm{MHz}$) since in the  rf-SQUID the flux
  noise decreases at higher  rf tank frequencies and the detector operates in a non-adiabatic regime where rf-SQUID
  amplifiers show lower flux noise \cite{ClarkeBook} ; measurements of excess noise (integrated NEP) as function of
  temperature since a weak dependency to $T^{2/3}$
  is expected \cite{Kurkij}; experiments with TESs with different
  weak-link parameters ($I_c$ and $R_N$) to achieve  a non-hysteretic regime.

\begin{acknowledgements}

We thank Manuela Popescu and Martijn Schoemans for their precious
technical help.

\end{acknowledgements}


\end{document}